\documentclass[a4paper,aps,prl,preprintnumbers,amsmath,amssymb,twocolumn,longbibliography]{revtex4-1} 

\usepackage{amsmath}
\usepackage{amssymb}
\usepackage{verbatim}
\usepackage{graphicx}
\usepackage{color}
\usepackage{xspace}
\usepackage{hyperref}

\begin{document}
 
\title{Quantum annealing with longitudinal bias fields}

\author{Tobias Gra{\ss}}
\affiliation{Joint Quantum Institute, NIST and University of Maryland, College Park, Maryland, 20742, USA}

\begin{abstract}
Quantum annealing aims at solving hard computational problems through adiabatic state preparation. Here, I propose to use inhomogeneous longitudinal magnetic fields to enhance the efficiency of the annealing. Such fields are able to bias the annealing dynamics into the desired solution, and in many cases, suitable field configurations can be found iteratively. Alternatively, the longitudinal fields can also be applied as an antibias which filters out unwanted contributions from the final state. This strategy is particularly well suited for instances which are difficult to solve within the standard quantum annealing approach. By numerically simulating the dynamics for small instances of the exact cover problem, the performance of these different strategies is investigated.
\end{abstract}

\maketitle

Thanks to the spectacular progress in controlling quantum systems, quantum dynamics has become a tool for solving hard computational problems. A strategy is quantum annealing \cite{das08,albash18,hauke19}: By incorporating the computational problem in the Hamiltonian, its solution is provided by the ground state which can be prepared by adiabatically reducing  quantum fluctuations. While early numerical studies have suggested that quantum annealing can reach even complicated ground states remarkably fast \cite{kadowaki98,farhi01}, later analysis indicated that during the evolution the system undergoes a first-order localization transition with exponentially small energy gaps \cite{joerg08,young10,joerg10,altshuler10}. As adiabatic theorems demand that the velocity of Hamiltonian changes scales with an inverse power of the energy gap, the evolution is condemned to be either infinitely slow or non-adiabatic. Therefore, recent research has focussed on strategies to avoid this bottleneck of quantum annealing. Improvements have been reported using inhomogeneous transverse fields \cite{dickson11,farhi11}, or other non-commutative terms which render the Hamiltonian non-stoquastic \cite{seki12,hormozi17,albash19,oxfidan19}. These tricks can turn the first-order phase transition into a second-order transition. Yet another strategy exploits the knowledge of an approximate solution, which can be incorporated in the initial configuration, and which can then be improved through reverse annealing \cite{Perdomo-Ortiz2011,ohkuwa18,baldwin18}. Similarly, combining state preparation and quantum annealing techniques gives rise to a variety of hybrid algorithms \cite{chancellor17,grass17}.

Here, I present a method which, similar to reverse annealing, exploits the knowledge of an approximate solution, see also Refs. \cite{duan13,ramezanpour17}. Concretely, a bias towards good solutions is incorporated by inhomogeneous longitudinal fields in the driving term. Simulating the dynamics for up to 15-bit instances of the exact cover problem, I show that the annealing success rate is greatly enhanced due to the bias. Specifically, when the error of the bias remains below 20\%, success rates $p>0.8$ can be achieved within times for which conventional annaeling yields only $p\sim 0.2$. Importantly, the bias can be adjusted iteratively such that for most instances convergence to the desired solution is achieved. Alternatively, by reversing the sign of the bias field, an ``anti-bias'' allows for reducing the probability of ending up in the wrong solution.

After describing the system ingredients required for these algorithms, this manuscript studies the annealing dynamics (i) in randomly generated bias fields, (ii) in iteratively generated bias fields, and (iii) in anti-bias fields. Significant performance enhancements are found if the bias is close to the optimal solution. Within the iterative scheme, this is typically not the case for the hardest instances, and in these cases the use of antibias fields provides a better strategy.

\textit{System.}
As in conventional quantum annealing, the dynamics is generated by a Hamiltonian of the form:
\begin{align}
 H(t) = A(t) H_{\rm p} + B(t) H_{\rm q}.
\end{align}
Here, $H_{\rm p}$ is the problem Hamiltonian, such that minimizing $\langle H_{\rm p} \rangle$ solves the computational problem. Typically, $H_{\rm p}$ is a classical spin model, and a huge variety of relevant optimization problems can be mapped onto Ising-type models, $H_{\rm p}=\sum_{mn} J_{mn} \sigma_m^z \sigma_n^z$, known to be of NP-hard computational complexity \cite{barahona}. The second term, $H_{\rm q}$, is chosen such that it does not commute with $H_{\rm p}$. Thereby, it introduces quantum fluctuations, and eigenstates of $H(t)$ are superpositions of different spin configurations. A common choice for $H_{\rm q}$ is a homogeneous transverse field, e.g. $H_{\rm q}=\sum_m \sigma_m^x$. The annealing schedule is controlled by time-dependent functions $A(t)$ and $B(t)$ chosen such that initially $B(0)\gg A(0)$ (or $A(0)=0$), whereas at the end ($t=T$), the opposite relation holds, $A(T) \gg B(T)$ (or $B(0)=0$).

The strategies proposed here exploit longitudinal field components within the quantum term $H_{\rm q}$:
\begin{align}
\label{Hq}
 H_{\rm q}= \sum_m (\sigma_m^x + h_m \sigma_m^z).
\end{align}
I refer to $h_m$ as the ``bias'' or ``anti-bias'' fields, as they will be used to incorporate some prior knowledge about the solution. As a bias, the fields shall force the annealing dynamics towards a certain configuration, whereas as an anti-bias, they may be used to avoid certain undesired outcomes.

I have simulated these protocols for random instances of the exact cover problem, an NP-hard optimization problem which has become paradigmatic for the study of quantum annealing algorithms (cf. Ref. \onlinecite{farhi01,altshuler10}). A problem instance is given by $N$ bits and $M$ clauses, each selecting a triple of bits. A clause is fulfilled if exactly two of the three bits take the value 1. Exact cover seeks a configuration which fulfills as many clauses as possible. In the language of a spin model, the problem is described by the Hamiltonian
\begin{align}
 H_{\rm p} = \sum_C \left(\sigma_{C(1)}^z + \sigma_{C(2)}^z +\sigma_{C(3)}^z -1 \right)^2 \equiv \sum_C h(C),
\end{align}
where the sum is over all clauses $C$, and $C(i)$ with $i=1,2,3$ denote the three spins/bits selected by clause $C$. Fulfillment of a clause leads to $\langle h(C) \rangle =0$, whereas its violation yields $\langle h(C) \rangle >0$. The existence of an assignment which satisfies all clauses corresponds to the ground state energy of $H_{\rm p}$ being zero. Analysis of the exact cover problem has suggested that hard-to-solve instances occur at a ratio of $N/M \approx 0.6$ \cite{kalapala08}, and instances with a unique satisfying assignment are amongst the hardest ones. In the following, I will only consider randomly generated instances with unique satisfying assignment.

For my simulations, I choose an exponential ramp, $B(t)=B(0) \exp(-t/\tau)$, and $A(t)={\rm const.}$. With $B(0)\gg 1$ (for concreteness: $B(0)=50$), the initial ground state matches the paramagnetic phase of $H_{\rm q}$. The annealing terminates at time $T=10\tau$, after convergence to a final state has been reached (typically at $t\approx 5\tau$). The exponential ramp is not only common to experimental adiabatic state preparation schemes (cf. Ref. \cite{islam13}), but it also enhances the performance of the annealing compared to a linear ramp. In order to test the performance of the bias field, I choose $\tau=1$ (i.e. on the order of inverse interactions strengths), well within typical coherence times of many quantum simulators. For the standard annealing protocol (i.e. without bias), this choice yields a success rate of $p\equiv |\langle \Psi(T) | S^0 \rangle |^2\approx 0.25$, depending on the problem size (varied from 10 to 15 spins). Here, $|\Psi(T)\rangle$ denotes the final state, and $|S^0\rangle$ the optimal solution.

\begin{figure}
\centering
\includegraphics[width=0.48\textwidth, angle=0]{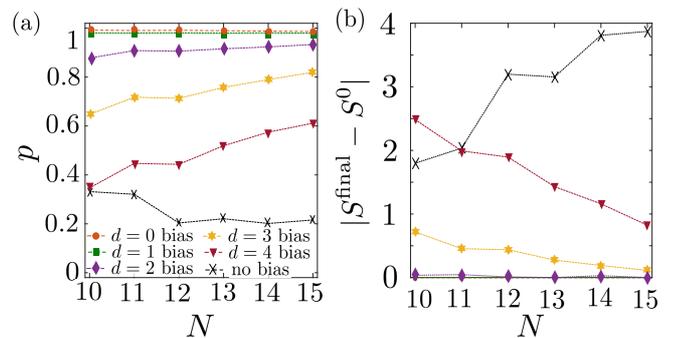}
\caption{
\label{fig1} {\bf Performance of quantum annealing in the presence of bias fields.} 
The plots show (a) the average success probability $p$, and (b) the average Hamming distance $|S^{\rm final}-S^0|$ of the final state from the global minimum, as a function of the problem size $N$. The different curves correspond to different bias fields, obtained from the optimal solution by randomly flipping $d$ spins. Averages are taken over 500 (200, 124) randomly generated instances of exact cover with unique satisfying assignment, for $N \leq 12$  ($N\leq 14$, $N=15$). The presence of a bias field is seen to enhance the performance of the annealing process: In contrast to the case without bias, the annealing outcome improves for larger $N$ in the presence of a bias.
}
\end{figure}

\textit{Annealing with bias field.}
Ideally, standard quantum annealing rotates a generic $x$-polarized state to a specific $z$-polarized state, but violation of the adiabatic condition leads to a superposition of various $z$-polarized states. The final weight of the optimal configuration can be increased by initial state preparation based on an approximate guess of the solution. Specifically, the fields in Eq.~(\ref{Hq}) allow for polarizing the initial state within the $xz$-plane.

Let $S\equiv\{s_1,\dots,s_N\}$ denote a spin configuration which ressembles the solution of the problem, denoted by $S^0\equiv \{s_1^0,\dots,s_N^0\}$. Here, ressemblance is measured by the Hamming distance $|S-S^0|$. By anti-aligning $h_m$ with $S$, i.e. $h_m=-s_m$, an energetic bias towards configurations with small Hamming distance from $S$ is provided. By including the bias field into $H_{\rm q}$, the adiabatic theorem still guarantees to reach the ground state of $H_{\rm p}$ when parameter changes are sufficiently slow.

Fig.~\ref{fig1} compares the performance of unbiased and biased quantum annealing, for different bias configurations $S$, characterized through their Hamming distance $d=|S-S^0|$. The figure of merit in Fig.~\ref{fig1}(a) is the success rate $p=\langle p_i \rangle$, averaged over each instance's individual success probability $p_i$.
The presence of a bias field greatly increases $p$ at any size ($10\leq N \leq 15$). Notably, for a bias with $d\leq1$, the success probability is almost unity. Importantly, larger problem instances can accomodate for larger $d$. For instance, while a bias field of $d=4$ does increase $p$ for $N=10$, an almost threefold improvement is obtained for $N=15$.

Alternatively, the fidelity can be measured in terms of observables, specifically by the spin configuration $S^{\rm final}$ obtained from the final spin expectation values: $s_m^{\rm final} = {\rm sign}(\langle \sigma_m^z(T) \rangle)$. The quality of $S^{\rm final}$ is characterized through its Hamming distance from the solution [see Fig.~\ref{fig1}(b)].  On average, the final Hamming distance drops below 1 for $N=15$ in the biased annealing even for $d=4$, constituting a more than fourfold improvement as compared to the unbiased process.

\begin{figure}
\centering
\includegraphics[width=0.48\textwidth, angle=0]{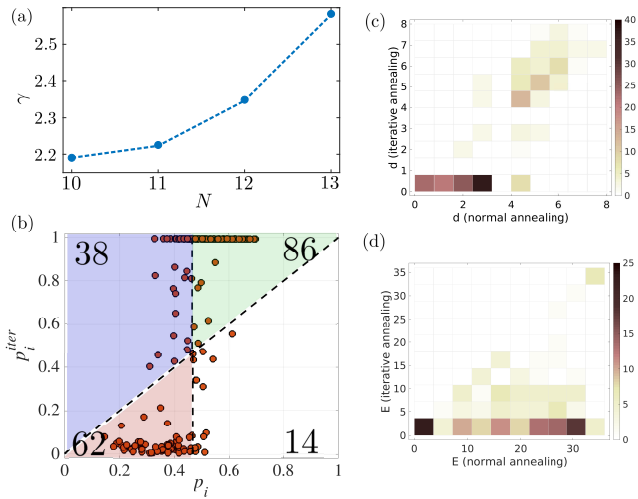}
\caption{
\label{fig2} {\bf Performance of iterative quantum annealing.} 
(a) Success enhancement $\gamma$ of the iterative scheme vs. problem size. (b) Individual success probabilities of 200 instances ($N=13$): success probability of the standard scheme is plotted versus success probability of the iterative scheme. A vertical and a diagonal line divide the diagram into four sectors, with populations given by the numbers. The iterative scheme outperforms the standard annealing for all instances above the diagonal line. The 100 hardest instances are left from the vertical line. (c) Contour histogram over Hamming distances from the desired solution for the outcome $S^{\rm final}$ of standard and iterative quantum annealing scheme. (d) Contour histogram over energy/cost function related to $S^{\rm final}$ in standard and iterative quantum annealing.
}
\end{figure}

\textit{Iterative quantum annealing.}
The data in Fig.~\ref{fig1} allows for the following observations: At all system sizes, unbiased annealing produces a final result with an average Hamming distance between 2 (for $N=10$) and 4 (for $N=15$). On the other hand, a bias field at that distance already yields improved results. This suggests an iterative algorithm, which starts with an unbiased run, and feeds subsequent runs with the outcome of the previous process, $s_m^{\rm final} = {\rm sign}(\langle \sigma_m^z \rangle^{\rm final})$. The algorithm stops when subsequent runs yield the same $s_m^{\rm final}$. The performance of this strategy is investigated below.

The average success probability, $p^{\rm iter}=\langle p_i^{\rm iter}\rangle$, as achieved at the end of such iterative process (which on average is between 2.6 steps for $N=10$ and 2.9 steps for $N=13$) is more than doubled compared to the success probability $p^{\rm st}$ in standard annealing. Strikingly, this enhancement $\gamma \equiv p^{\rm iter}/p^{\rm st}$ increases with the problem size, see Fig.~\ref{fig2}(a). Other figures of merit show improvements as well: The mean energy at the end of the iterative schedule is found between 3.1 (for $N=10$) and 4.6 (for $N=13$), significantly below the corresponding value for standard quantum annealing (between 12.3 and 19.3). 

However, the iterative scheme also leads to an extremely large standard deviation in the success probability, e.g. $p^{\rm iter}=0.57 \pm 0.46$ for $N=13$ (compared to $p^{\rm st}=0.22\pm 0.10$). This indicates that, in contrast to the standard protocol which often ends up in a superposition of several classical states (one of which being the seeked solution), the iterative scheme tends towards a unique classical state. Then, $p^{\rm iter}$ is close to 1 when this state is the seeked solution, or  close to 0 in the opposite case. This splitting is illustrated in Fig.~\ref{fig2}(b), where the $p^{\rm iter}_i$ are plotted versus the corresponding $p^{\rm st}_i$ for 200 instances at $N=13$.

The success of iterative annealing strongly depends on the quality of the spin configuration $S^{\rm final}$ obtained after the first unbiased run. This is illustrated by the contour histogram of the final Hamming distance for standard and iterative annealing, shown in Fig.~\ref{fig2}(c). When standard annealing leads to a small Hamming distance, the iterative annealing is extremely likely to produce the correct solution. For 114 instances, the iterative $S^{\rm final}$ matched the correct solution (compared to only 23 matches with the standard protocol). 

However, for the hardest instances (with low success probability $p_i$), the outcome of the unbiased first run usually does not provide a suitable bias. The iterative scheme then tends to a suboptimal solutions. This is seen from Fig.~\ref{fig2}(d), plotting the cost function value of the final configuration. Notably, the iterative scheme never leads to an increase of the cost function, and even among the hardest instances [in the red-shaded area in Fig.~\ref{fig2}(b)], the average cost function of the interative scheme is 7.9 compared to 20.5 in standard quantum annealing. In 154 (of 200) cases, the iterative scheme yields a lower cost function than standard quantum annealing.

In summary, for most instances iterative annealing yields the best solution with very high fidelity, but for the particularly interesting class of instances which are hard for standard quantum annealing, it only tends to suboptimal solutions. Possibly, by exploiting additional information from other methods, e.g. from classical annealing methods \cite{kirkpatrick83}, improvements on the iterative scheme can be achieved. In the following, though, I will discuss another strategy which in some sense is opposite to the approach discussed so far. 

\begin{table}
 \begin{tabular}{|l||c|c|c|c|c|c|c|c|c|}
\hline
$N$ & $\tau^{\rm ab}$ & $\tau^{\rm st}$ & $\bar p$ & $p^{\rm f}$ & $p$  & $\bar p(5\%)$ & $p^{\rm f}(5\%)$ & $p(5\%)$ \\
\hline
\hline
11 & 3.3 & 3.8  & 0.33 & 0.37 & 0.31 & 0.18 & 0.30 & 0.11 \\
\hline
12 & 4.2 & 6.8  & 0.23 & 0.28 & 0.20 & 0.11 & 0.21 & 0.046 \\
\hline
 \end{tabular}
\caption{ \label{table} The performance of the annealing with an antibias field is quantified by the average number of steps, $\tau$, and by different measures for the success rate (defined in the text). The algorithm performs well among the $5\%$ of instances which are hardest for the standard quantum annealing protocol.}
\end{table}

\begin{figure}
\centering
\includegraphics[width=0.49\textwidth, angle=0]{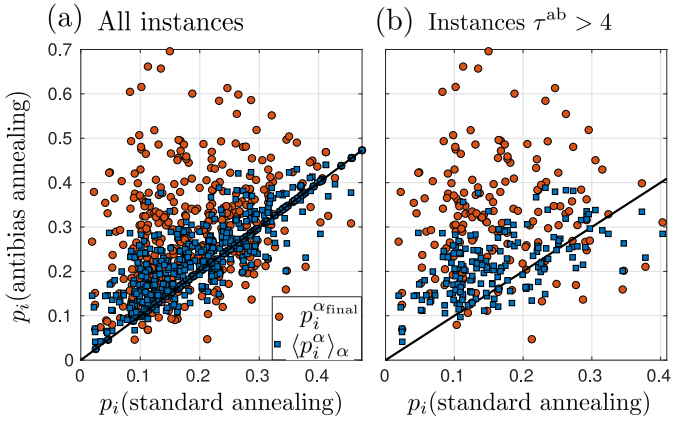}
\caption{
\label{fig3} {\bf Comparison of antibias and standard quantum annealing.} 
The individual success probabilities $p_i$ of standard annealing are plotted vs. their success probabilities in the presence of an antibias field, measured either by the average over all iteration steps, $\langle p_i^{\alpha} \rangle_\alpha$, or by the final success probability $p_i^{\alpha_{\rm final}}$ after the last step. Panel (a) shows the data for all 500 instances considered at $N=12$, whereas panel (b) considers only those instance which took more than 4 annealing steps ($\tau^{\rm ab}>4$). The majority of data points lies above the solid line, indicating that for these instances the antibias scheme has outperformed the standard annealing scheme.
}
\end{figure}

\textit{Annealing with anti-bias fields.}
As seen above, for the hardest instances standard annealing does not produce a suitable guess for the iterative scheme, and thus, a bias field would drive the annealer away from the desired solution. This motivates an alternative strategy: For hard instances, it might actually be beneficial to use the outcome $s_m$ of an annealing run as an anti-bias field, e.g. $h_m=s_m$ instead of $h_m=-s_m$. Such field will then act as a filter for an unwanted spin configuration, thereby enhancing the chance of finding the correct state. 

Since \textit{a priori} the quality of the outcome is unknown, the antibias fields must be chosen sufficiently weak to avoid detrimental effects when the initial outcome is good. Moreover, the filtering becomes most efficient, if the algorithm keeps memory of all previously obtained configurations. Thus, the antibias field should be accumulative. In addition, rather than deriving the antibias field from the final spin expectation values, one may instead use a single projective measurement of $\sigma_m^z$, which tends to have a larger Hamming distance from the correct solution than the expectation values. In this way, the algorithm becomes non-deterministic, but importantly, all overhead associated with the measurement of expectation values is removed. 

 Taking these considerations into accout, I define the antibias as $h_m \equiv h \sum_{\alpha} s_m^\alpha$, with $h<1$ being a constant, and $\{s_m^\alpha\}=s_1^\alpha,\dots,s_N^\alpha$ denoting the spin configurations obtained from a projective measurement after the $\alpha$th run. For concreteness, in the results presented below, I have chosen $h=0.1$. As for standard annealing, but in contrast to the iterative scheme discussed earlier, there is no self-termination of the algorithm, and the simulation stops when the $E=0$ solution was obtained. Due to the projective measurements, this procedure involves a non-determinisitic number of steps.

As the algorithm keeps running, the antibias field will accumulate on those spins which repeatedly take the same value. Thus, if there are a few spins which distinguish the optimal solution from a larger pool of suboptimal solutions, the algorithm tends towards filtering out these solutions. The overall performance of the algorithm can be quantified by the number $\tau$ of steps until the optimal solution is found. For standard annealing, $\tau^{\rm st}= \langle p_i^{-1} \rangle$. In the presence of an accumulative antibias field, success probabilities $p_i^\alpha$ indeterministically change from run to run (denoted by $\alpha$), and an overall average $\tau^{\rm ab}$ is obtained from sampling over 500 different instances. For the largest problem size considered ($N=12$), $\tau^{\rm ab}$ is considerably lower than $\tau^{\rm st}$, see Table~\ref{table}.

For a performance benchmark in terms of success probabilities, I have consider the success probability $p_i^\alpha$ for instance $i$ after run $\alpha$, and average over all runs and over all instances, $\bar p \equiv \langle \langle p_i^{\alpha} \rangle_\alpha\rangle_i$. To better reflect the high success probabilities reached towards the end, I also consider the average success probability after the final run $\alpha_{\rm final}$, $p^{\rm f}\equiv \langle p_i^{\alpha_{\rm final}}\rangle_i$. Table~\ref{table} compares these quantities with the usual success rate $p$ in standard annealing. Averaged over all instances, the improvements seem marginal. However, the alorithm is designed to outperform standard annealing in the case of particularly hard instance, so the assessment changes when, for instance, only the hardest $5\%$ of instances (i.e. 25 instances) are considered, denoted $\bar p(5\%)$, $p^{\rm f}(5\%)$, and $p(5\%)$. As also seen from Table~\ref{table}, the antibias fields then show a significant effect: While for $N=12$ the standard annealing success rate remains below 0.05, for antibias annealing $p^{\rm f}(5\%)$ is above 0.2.

This observation is also illustrated in Fig.~\ref{fig3}, plotting the individual success rates of standard annealing $p_i$ vs. $\langle p_i^{\alpha} \rangle_\alpha$ and vs. $p_i^{\alpha_{\rm final}}$. In panel (a), data for all 500 instances is plotted, while panel (b) focuses on those (typically hard) instances where several annealing steps were needed ($\tau^{\rm ab}>4$). In both cases, roughly 80\% of the data points show an increase of success probability due to the antibias field, but the enhancement is seen to be more pronounced for the instances shown in (b).

\textit{Outlook and summary.}
This paper proposes to use longitudinal fields to either drive the annealing dynamics into a desired state, if a good guess is available, or alternatively, to filter out undesired configurations. For small instances of the exact cover problem, both strategies have been shown to produce performance enhancement compared to standard quantum annealing. For the most efficient exploration of these possibilities, it will be necessary to go beyond the costly classical simulation of the dynamics, and to implement the proposed schemes in a quantum device. Currently, the state-of-art platform for quantum annealing is the D-Wave machine, a commercial device built of 2048 superconducting qubits. Notably, inhomogeneous longitudinal fields as needed for the strategies presented here have been implemented in recent experiments with the D-Wave \cite{adame18}, but the required independent scheduling of $zz$-interactions and $z$-fields might currently not be available. In the future, various atomic platforms may also become available for quantum annealing purposes, such as trapped ions \cite{hauke15,grass16}, Rydberg atoms \cite{glaetzle17}, or atoms in optical cavities \cite{torggler17}. In these systems, magnetic fields can be achieved and freely tuned through AC Stark shifts with single-atom resolution, and both longitudinal and transverse fields are a standard ingredient to many atomic quantum simulations, e.g. see Ref. \cite{richerme13}. The initial state which is polarized in the $xz$-plane can be achieved either by reverse annealing, i.e. by starting with a longitudinal bias field and adiabatically switching on the homogeneous transverse field, or by starting from the fully polarized $\sigma^x$ state as in standard quantum annealing, and subsequently acting with single-qubit gates performing a $\pm \pi/4$ rotation.

\begin{acknowledgments}
The author acknowledges the University of Maryland supercomputing resources (\url{http://hpcc.umd.edu}) made available for conducting the research reported in this paper.
\end{acknowledgments}


\begin{thebibliography}{32}%
\makeatletter
\providecommand \@ifxundefined [1]{%
 \@ifx{#1\undefined}
}%
\providecommand \@ifnum [1]{%
 \ifnum #1\expandafter \@firstoftwo
 \else \expandafter \@secondoftwo
 \fi
}%
\providecommand \@ifx [1]{%
 \ifx #1\expandafter \@firstoftwo
 \else \expandafter \@secondoftwo
 \fi
}%
\providecommand \natexlab [1]{#1}%
\providecommand \enquote  [1]{``#1''}%
\providecommand \bibnamefont  [1]{#1}%
\providecommand \bibfnamefont [1]{#1}%
\providecommand \citenamefont [1]{#1}%
\providecommand \href@noop [0]{\@secondoftwo}%
\providecommand \href [0]{\begingroup \@sanitize@url \@href}%
\providecommand \@href[1]{\@@startlink{#1}\@@href}%
\providecommand \@@href[1]{\endgroup#1\@@endlink}%
\providecommand \@sanitize@url [0]{\catcode `\\12\catcode `\$12\catcode
  `\&12\catcode `\#12\catcode `\^12\catcode `\_12\catcode `\%12\relax}%
\providecommand \@@startlink[1]{}%
\providecommand \@@endlink[0]{}%
\providecommand \url  [0]{\begingroup\@sanitize@url \@url }%
\providecommand \@url [1]{\endgroup\@href {#1}{\urlprefix }}%
\providecommand \urlprefix  [0]{URL }%
\providecommand \Eprint [0]{\href }%
\providecommand \doibase [0]{http://dx.doi.org/}%
\providecommand \selectlanguage [0]{\@gobble}%
\providecommand \bibinfo  [0]{\@secondoftwo}%
\providecommand \bibfield  [0]{\@secondoftwo}%
\providecommand \translation [1]{[#1]}%
\providecommand \BibitemOpen [0]{}%
\providecommand \bibitemStop [0]{}%
\providecommand \bibitemNoStop [0]{.\EOS\space}%
\providecommand \EOS [0]{\spacefactor3000\relax}%
\providecommand \BibitemShut  [1]{\csname bibitem#1\endcsname}%
\let\auto@bib@innerbib\@empty
\bibitem [{\citenamefont {Das}\ and\ \citenamefont
  {Chakrabarti}(2008)}]{das08}%
  \BibitemOpen
  \bibfield  {author} {\bibinfo {author} {\bibfnamefont {Arnab}\ \bibnamefont
  {Das}}\ and\ \bibinfo {author} {\bibfnamefont {Bikas~K.}\ \bibnamefont
  {Chakrabarti}},\ }\bibfield  {title} {\enquote {\bibinfo {title}
  {{Colloquium: Quantum annealing and analog quantum computation}},}\ }\href
  {\doibase 10.1103/RevModPhys.80.1061} {\bibfield  {journal} {\bibinfo
  {journal} {Rev. Mod. Phys.}\ }\textbf {\bibinfo {volume} {80}},\ \bibinfo
  {pages} {1061--1081} (\bibinfo {year} {2008})}\BibitemShut {NoStop}%
\bibitem [{\citenamefont {Albash}\ and\ \citenamefont
  {Lidar}(2018)}]{albash18}%
  \BibitemOpen
  \bibfield  {author} {\bibinfo {author} {\bibfnamefont {Tameem}\ \bibnamefont
  {Albash}}\ and\ \bibinfo {author} {\bibfnamefont {Daniel~A.}\ \bibnamefont
  {Lidar}},\ }\bibfield  {title} {\enquote {\bibinfo {title} {{Adiabatic
  quantum computation}},}\ }\href {\doibase 10.1103/RevModPhys.90.015002}
  {\bibfield  {journal} {\bibinfo  {journal} {Rev. Mod. Phys.}\ }\textbf
  {\bibinfo {volume} {90}},\ \bibinfo {pages} {015002} (\bibinfo {year}
  {2018})}\BibitemShut {NoStop}%
\bibitem [{\citenamefont {Hauke}\ \emph {et~al.}(2019)\citenamefont {Hauke},
  \citenamefont {Katzgraber}, \citenamefont {Lechner}, \citenamefont
  {Nishimori},\ and\ \citenamefont {Oliver}}]{hauke19}%
  \BibitemOpen
  \bibfield  {author} {\bibinfo {author} {\bibfnamefont {Philipp}\ \bibnamefont
  {Hauke}}, \bibinfo {author} {\bibfnamefont {Helmut~G.}\ \bibnamefont
  {Katzgraber}}, \bibinfo {author} {\bibfnamefont {Wolfgang}\ \bibnamefont
  {Lechner}}, \bibinfo {author} {\bibfnamefont {Hidetoshi}\ \bibnamefont
  {Nishimori}}, \ and\ \bibinfo {author} {\bibfnamefont {William~D.}\
  \bibnamefont {Oliver}},\ }\bibfield  {title} {\enquote {\bibinfo {title}
  {{Perspectives of quantum annealing: Methods and implementations}},}\
  }\href@noop {} {\bibfield  {journal} {\bibinfo  {journal} {arXiv 1903.06559}\
  } (\bibinfo {year} {2019})}\BibitemShut {NoStop}%
\bibitem [{\citenamefont {Kadowaki}\ and\ \citenamefont
  {Nishimori}(1998)}]{kadowaki98}%
  \BibitemOpen
  \bibfield  {author} {\bibinfo {author} {\bibfnamefont {Tadashi}\ \bibnamefont
  {Kadowaki}}\ and\ \bibinfo {author} {\bibfnamefont {Hidetoshi}\ \bibnamefont
  {Nishimori}},\ }\bibfield  {title} {\enquote {\bibinfo {title} {{Quantum
  annealing in the transverse Ising model}},}\ }\href {\doibase
  10.1103/PhysRevE.58.5355} {\bibfield  {journal} {\bibinfo  {journal} {Phys.
  Rev. E}\ }\textbf {\bibinfo {volume} {58}},\ \bibinfo {pages} {5355--5363}
  (\bibinfo {year} {1998})}\BibitemShut {NoStop}%
\bibitem [{\citenamefont {Farhi}\ \emph {et~al.}(2001)\citenamefont {Farhi},
  \citenamefont {Goldstone}, \citenamefont {Gutmann}, \citenamefont {Lapan},
  \citenamefont {Lundgren},\ and\ \citenamefont {Preda}}]{farhi01}%
  \BibitemOpen
  \bibfield  {author} {\bibinfo {author} {\bibfnamefont {Edward}\ \bibnamefont
  {Farhi}}, \bibinfo {author} {\bibfnamefont {Jeffrey}\ \bibnamefont
  {Goldstone}}, \bibinfo {author} {\bibfnamefont {Sam}\ \bibnamefont
  {Gutmann}}, \bibinfo {author} {\bibfnamefont {Joshua}\ \bibnamefont {Lapan}},
  \bibinfo {author} {\bibfnamefont {Andrew}\ \bibnamefont {Lundgren}}, \ and\
  \bibinfo {author} {\bibfnamefont {Daniel}\ \bibnamefont {Preda}},\ }\bibfield
   {title} {\enquote {\bibinfo {title} {{A Quantum Adiabatic Evolution
  Algorithm Applied to Random Instances of an NP-Complete Problem}},}\ }\href
  {\doibase 10.1126/science.1057726} {\bibfield  {journal} {\bibinfo  {journal}
  {Science}\ }\textbf {\bibinfo {volume} {292}},\ \bibinfo {pages} {472--475}
  (\bibinfo {year} {2001})}\BibitemShut {NoStop}%
\bibitem [{\citenamefont {J{\"o}rg}\ \emph {et~al.}(2008)\citenamefont
  {J{\"o}rg}, \citenamefont {Krzakala}, \citenamefont {Kurchan},\ and\
  \citenamefont {Maggs}}]{joerg08}%
  \BibitemOpen
  \bibfield  {author} {\bibinfo {author} {\bibfnamefont {Thomas}\ \bibnamefont
  {J{\"o}rg}}, \bibinfo {author} {\bibfnamefont {Florent}\ \bibnamefont
  {Krzakala}}, \bibinfo {author} {\bibfnamefont {Jorge}\ \bibnamefont
  {Kurchan}}, \ and\ \bibinfo {author} {\bibfnamefont {A.~C.}\ \bibnamefont
  {Maggs}},\ }\bibfield  {title} {\enquote {\bibinfo {title} {{Simple Glass
  Models and Their Quantum Annealing}},}\ }\href {\doibase
  10.1103/PhysRevLett.101.147204} {\bibfield  {journal} {\bibinfo  {journal}
  {Phys. Rev. Lett.}\ }\textbf {\bibinfo {volume} {101}},\ \bibinfo {pages}
  {147204} (\bibinfo {year} {2008})}\BibitemShut {NoStop}%
\bibitem [{\citenamefont {Young}\ \emph {et~al.}(2010)\citenamefont {Young},
  \citenamefont {Knysh},\ and\ \citenamefont {Smelyanskiy}}]{young10}%
  \BibitemOpen
  \bibfield  {author} {\bibinfo {author} {\bibfnamefont {A.~P.}\ \bibnamefont
  {Young}}, \bibinfo {author} {\bibfnamefont {S.}~\bibnamefont {Knysh}}, \ and\
  \bibinfo {author} {\bibfnamefont {V.~N.}\ \bibnamefont {Smelyanskiy}},\
  }\bibfield  {title} {\enquote {\bibinfo {title} {{First-Order Phase
  Transition in the Quantum Adiabatic Algorithm}},}\ }\href {\doibase
  10.1103/PhysRevLett.104.020502} {\bibfield  {journal} {\bibinfo  {journal}
  {Phys. Rev. Lett.}\ }\textbf {\bibinfo {volume} {104}},\ \bibinfo {pages}
  {020502} (\bibinfo {year} {2010})}\BibitemShut {NoStop}%
\bibitem [{\citenamefont {J{\"o}rg}\ \emph {et~al.}(2010)\citenamefont
  {J{\"o}rg}, \citenamefont {Krzakala}, \citenamefont {Semerjian},\ and\
  \citenamefont {Zamponi}}]{joerg10}%
  \BibitemOpen
  \bibfield  {author} {\bibinfo {author} {\bibfnamefont {Thomas}\ \bibnamefont
  {J{\"o}rg}}, \bibinfo {author} {\bibfnamefont {Florent}\ \bibnamefont
  {Krzakala}}, \bibinfo {author} {\bibfnamefont {Guilhem}\ \bibnamefont
  {Semerjian}}, \ and\ \bibinfo {author} {\bibfnamefont {Francesco}\
  \bibnamefont {Zamponi}},\ }\bibfield  {title} {\enquote {\bibinfo {title}
  {{First-Order Transitions and the Performance of Quantum Algorithms in Random
  Optimization Problems}},}\ }\href {\doibase 10.1103/PhysRevLett.104.207206}
  {\bibfield  {journal} {\bibinfo  {journal} {Phys. Rev. Lett.}\ }\textbf
  {\bibinfo {volume} {104}},\ \bibinfo {pages} {207206} (\bibinfo {year}
  {2010})}\BibitemShut {NoStop}%
\bibitem [{\citenamefont {Altshuler}\ \emph {et~al.}(2010)\citenamefont
  {Altshuler}, \citenamefont {Krovi},\ and\ \citenamefont
  {Roland}}]{altshuler10}%
  \BibitemOpen
  \bibfield  {author} {\bibinfo {author} {\bibfnamefont {Boris}\ \bibnamefont
  {Altshuler}}, \bibinfo {author} {\bibfnamefont {Hari}\ \bibnamefont {Krovi}},
  \ and\ \bibinfo {author} {\bibfnamefont {J{\'e}r{\'e}mie}\ \bibnamefont
  {Roland}},\ }\bibfield  {title} {\enquote {\bibinfo {title} {{Anderson
  localization makes adiabatic quantum optimization fail}},}\ }\href {\doibase
  10.1073/pnas.1002116107} {\bibfield  {journal} {\bibinfo  {journal} {Proc.
  Nat. Acad. Sci. USA}\ }\textbf {\bibinfo {volume} {107}},\ \bibinfo {pages}
  {12446--12450} (\bibinfo {year} {2010})}\BibitemShut {NoStop}%
\bibitem [{\citenamefont {Dickson}\ and\ \citenamefont
  {Amin}(2011)}]{dickson11}%
  \BibitemOpen
  \bibfield  {author} {\bibinfo {author} {\bibfnamefont {Neil~G.}\ \bibnamefont
  {Dickson}}\ and\ \bibinfo {author} {\bibfnamefont {M.~H.~S.}\ \bibnamefont
  {Amin}},\ }\bibfield  {title} {\enquote {\bibinfo {title} {{Does Adiabatic
  Quantum Optimization Fail for NP-Complete Problems?}}}\ }\href {\doibase
  10.1103/PhysRevLett.106.050502} {\bibfield  {journal} {\bibinfo  {journal}
  {Phys. Rev. Lett.}\ }\textbf {\bibinfo {volume} {106}},\ \bibinfo {pages}
  {050502} (\bibinfo {year} {2011})}\BibitemShut {NoStop}%
\bibitem [{\citenamefont {Farhi}\ \emph {et~al.}(2011)\citenamefont {Farhi},
  \citenamefont {Goldstone}, \citenamefont {Gosset}, \citenamefont {Gutmann},
  \citenamefont {Meyer},\ and\ \citenamefont {Shor}}]{farhi11}%
  \BibitemOpen
  \bibfield  {author} {\bibinfo {author} {\bibfnamefont {Edward}\ \bibnamefont
  {Farhi}}, \bibinfo {author} {\bibfnamefont {Jeffrey}\ \bibnamefont
  {Goldstone}}, \bibinfo {author} {\bibfnamefont {David}\ \bibnamefont
  {Gosset}}, \bibinfo {author} {\bibfnamefont {Sam}\ \bibnamefont {Gutmann}},
  \bibinfo {author} {\bibfnamefont {Harvey~B.}\ \bibnamefont {Meyer}}, \ and\
  \bibinfo {author} {\bibfnamefont {Peter}\ \bibnamefont {Shor}},\ }\bibfield
  {title} {\enquote {\bibinfo {title} {{Quantum adiabatic algorithms, small
  gaps, and different paths}},}\ }\href@noop {} {\bibfield  {journal} {\bibinfo
   {journal} {Quant. Inf. Comput.}\ }\textbf {\bibinfo {volume} {11}},\
  \bibinfo {pages} {181} (\bibinfo {year} {2011})}\BibitemShut {NoStop}%
\bibitem [{\citenamefont {Seki}\ and\ \citenamefont
  {Nishimori}(2012)}]{seki12}%
  \BibitemOpen
  \bibfield  {author} {\bibinfo {author} {\bibfnamefont {Yuya}\ \bibnamefont
  {Seki}}\ and\ \bibinfo {author} {\bibfnamefont {Hidetoshi}\ \bibnamefont
  {Nishimori}},\ }\bibfield  {title} {\enquote {\bibinfo {title} {{Quantum
  annealing with antiferromagnetic fluctuations}},}\ }\href {\doibase
  10.1103/PhysRevE.85.051112} {\bibfield  {journal} {\bibinfo  {journal} {Phys.
  Rev. E}\ }\textbf {\bibinfo {volume} {85}},\ \bibinfo {pages} {051112}
  (\bibinfo {year} {2012})}\BibitemShut {NoStop}%
\bibitem [{\citenamefont {Hormozi}\ \emph {et~al.}(2017)\citenamefont
  {Hormozi}, \citenamefont {Brown}, \citenamefont {Carleo},\ and\ \citenamefont
  {Troyer}}]{hormozi17}%
  \BibitemOpen
  \bibfield  {author} {\bibinfo {author} {\bibfnamefont {Layla}\ \bibnamefont
  {Hormozi}}, \bibinfo {author} {\bibfnamefont {Ethan~W.}\ \bibnamefont
  {Brown}}, \bibinfo {author} {\bibfnamefont {Giuseppe}\ \bibnamefont
  {Carleo}}, \ and\ \bibinfo {author} {\bibfnamefont {Matthias}\ \bibnamefont
  {Troyer}},\ }\bibfield  {title} {\enquote {\bibinfo {title} {{Nonstoquastic
  Hamiltonians and quantum annealing of an Ising spin glass}},}\ }\href
  {\doibase 10.1103/PhysRevB.95.184416} {\bibfield  {journal} {\bibinfo
  {journal} {Phys. Rev. B}\ }\textbf {\bibinfo {volume} {95}},\ \bibinfo
  {pages} {184416} (\bibinfo {year} {2017})}\BibitemShut {NoStop}%
\bibitem [{\citenamefont {Albash}(2019)}]{albash19}%
  \BibitemOpen
  \bibfield  {author} {\bibinfo {author} {\bibfnamefont {Tameem}\ \bibnamefont
  {Albash}},\ }\bibfield  {title} {\enquote {\bibinfo {title} {{Role of
  nonstoquastic catalysts in quantum adiabatic optimization}},}\ }\href
  {\doibase 10.1103/PhysRevA.99.042334} {\bibfield  {journal} {\bibinfo
  {journal} {Phys. Rev. A}\ }\textbf {\bibinfo {volume} {99}},\ \bibinfo
  {pages} {042334} (\bibinfo {year} {2019})}\BibitemShut {NoStop}%
\bibitem [{\citenamefont {{I. Ozfidan \textit{et al.}}}(2019)}]{oxfidan19}%
  \BibitemOpen
  \bibfield  {author} {\bibinfo {author} {\bibnamefont {{I. Ozfidan \textit{et
  al.}}}},\ }\bibfield  {title} {\enquote {\bibinfo {title} {{Demonstration of
  nonstoquastic Hamiltonian in coupled superconducting flux qubits}},}\
  }\href@noop {} {\bibfield  {journal} {\bibinfo  {journal} {arXiv 1903.06139}\
  } (\bibinfo {year} {2019})}\BibitemShut {NoStop}%
\bibitem [{\citenamefont {Perdomo-Ortiz}\ \emph {et~al.}(2011)\citenamefont
  {Perdomo-Ortiz}, \citenamefont {Venegas-Andraca},\ and\ \citenamefont
  {Aspuru-Guzik}}]{Perdomo-Ortiz2011}%
  \BibitemOpen
  \bibfield  {author} {\bibinfo {author} {\bibfnamefont {Alejandro}\
  \bibnamefont {Perdomo-Ortiz}}, \bibinfo {author} {\bibfnamefont
  {Salvador~E.}\ \bibnamefont {Venegas-Andraca}}, \ and\ \bibinfo {author}
  {\bibfnamefont {Al{\'a}n}\ \bibnamefont {Aspuru-Guzik}},\ }\bibfield  {title}
  {\enquote {\bibinfo {title} {{A study of heuristic guesses for adiabatic
  quantum computation}},}\ }\href {\doibase 10.1007/s11128-010-0168-z}
  {\bibfield  {journal} {\bibinfo  {journal} {Quantum Information Processing}\
  }\textbf {\bibinfo {volume} {10}},\ \bibinfo {pages} {33--52} (\bibinfo
  {year} {2011})}\BibitemShut {NoStop}%
\bibitem [{\citenamefont {Ohkuwa}\ \emph {et~al.}(2018)\citenamefont {Ohkuwa},
  \citenamefont {Nishimori},\ and\ \citenamefont {Lidar}}]{ohkuwa18}%
  \BibitemOpen
  \bibfield  {author} {\bibinfo {author} {\bibfnamefont {Masaki}\ \bibnamefont
  {Ohkuwa}}, \bibinfo {author} {\bibfnamefont {Hidetoshi}\ \bibnamefont
  {Nishimori}}, \ and\ \bibinfo {author} {\bibfnamefont {Daniel~A.}\
  \bibnamefont {Lidar}},\ }\bibfield  {title} {\enquote {\bibinfo {title}
  {{Reverse annealing for the fully connected $p$-spin model}},}\ }\href
  {\doibase 10.1103/PhysRevA.98.022314} {\bibfield  {journal} {\bibinfo
  {journal} {Phys. Rev. A}\ }\textbf {\bibinfo {volume} {98}},\ \bibinfo
  {pages} {022314} (\bibinfo {year} {2018})}\BibitemShut {NoStop}%
\bibitem [{\citenamefont {Baldwin}\ and\ \citenamefont
  {Laumann}(2018)}]{baldwin18}%
  \BibitemOpen
  \bibfield  {author} {\bibinfo {author} {\bibfnamefont {C.~L.}\ \bibnamefont
  {Baldwin}}\ and\ \bibinfo {author} {\bibfnamefont {C.~R.}\ \bibnamefont
  {Laumann}},\ }\bibfield  {title} {\enquote {\bibinfo {title} {{Quantum
  algorithm for energy matching in hard optimization problems}},}\ }\href
  {\doibase 10.1103/PhysRevB.97.224201} {\bibfield  {journal} {\bibinfo
  {journal} {Phys. Rev. B}\ }\textbf {\bibinfo {volume} {97}},\ \bibinfo
  {pages} {224201} (\bibinfo {year} {2018})}\BibitemShut {NoStop}%
\bibitem [{\citenamefont {Chancellor}(2017)}]{chancellor17}%
  \BibitemOpen
  \bibfield  {author} {\bibinfo {author} {\bibfnamefont {Nicholas}\
  \bibnamefont {Chancellor}},\ }\bibfield  {title} {\enquote {\bibinfo {title}
  {{Modernizing quantum annealing using local searches}},}\ }\href {\doibase
  10.1088/1367-2630/aa59c4} {\bibfield  {journal} {\bibinfo  {journal} {New J.
  Phys.}\ }\textbf {\bibinfo {volume} {19}},\ \bibinfo {pages} {023024}
  (\bibinfo {year} {2017})}\BibitemShut {NoStop}%
\bibitem [{\citenamefont {Gra{\ss}}\ and\ \citenamefont
  {Lewenstein}(2017)}]{grass17}%
  \BibitemOpen
  \bibfield  {author} {\bibinfo {author} {\bibfnamefont {Tobias}\ \bibnamefont
  {Gra{\ss}}}\ and\ \bibinfo {author} {\bibfnamefont {Maciej}\ \bibnamefont
  {Lewenstein}},\ }\bibfield  {title} {\enquote {\bibinfo {title} {{Hybrid
  annealing: Coupling a quantum simulator to a classical computer}},}\ }\href
  {\doibase 10.1103/PhysRevA.95.052309} {\bibfield  {journal} {\bibinfo
  {journal} {Phys. Rev. A}\ }\textbf {\bibinfo {volume} {95}},\ \bibinfo
  {pages} {052309} (\bibinfo {year} {2017})}\BibitemShut {NoStop}%
\bibitem [{\citenamefont {Duan}\ \emph {et~al.}(2013)\citenamefont {Duan},
  \citenamefont {Zhang}, \citenamefont {Wu},\ and\ \citenamefont
  {Chen}}]{duan13}%
  \BibitemOpen
  \bibfield  {author} {\bibinfo {author} {\bibfnamefont {Qian-Heng}\
  \bibnamefont {Duan}}, \bibinfo {author} {\bibfnamefont {Shuo}\ \bibnamefont
  {Zhang}}, \bibinfo {author} {\bibfnamefont {Wei}\ \bibnamefont {Wu}}, \ and\
  \bibinfo {author} {\bibfnamefont {Ping-Xing}\ \bibnamefont {Chen}},\
  }\bibfield  {title} {\enquote {\bibinfo {title} {{An Alternative Approach to
  Construct the Initial Hamiltonian of the Adiabatic Quantum Computation}},}\
  }\href {\doibase 10.1088/0256-307x/30/1/010302} {\bibfield  {journal}
  {\bibinfo  {journal} {Chin. Phys. Lett.}\ }\textbf {\bibinfo {volume} {30}},\
  \bibinfo {pages} {010302} (\bibinfo {year} {2013})}\BibitemShut {NoStop}%
\bibitem [{\citenamefont {Ramezanpour}(2017)}]{ramezanpour17}%
  \BibitemOpen
  \bibfield  {author} {\bibinfo {author} {\bibfnamefont {A.}~\bibnamefont
  {Ramezanpour}},\ }\bibfield  {title} {\enquote {\bibinfo {title}
  {{Optimization by a quantum reinforcement algorithm}},}\ }\href {\doibase
  10.1103/PhysRevA.96.052307} {\bibfield  {journal} {\bibinfo  {journal} {Phys.
  Rev. A}\ }\textbf {\bibinfo {volume} {96}},\ \bibinfo {pages} {052307}
  (\bibinfo {year} {2017})}\BibitemShut {NoStop}%
\bibitem [{\citenamefont {Barahona}(1982)}]{barahona}%
  \BibitemOpen
  \bibfield  {author} {\bibinfo {author} {\bibfnamefont {F}~\bibnamefont
  {Barahona}},\ }\bibfield  {title} {\enquote {\bibinfo {title} {{On the
  computational complexity of Ising spin glass models}},}\ }\href
  {http://stacks.iop.org/0305-4470/15/i=10/a=028} {\bibfield  {journal}
  {\bibinfo  {journal} {J. Phys. A}\ }\textbf {\bibinfo {volume} {15}},\
  \bibinfo {pages} {3241} (\bibinfo {year} {1982})}\BibitemShut {NoStop}%
\bibitem [{\citenamefont {Kalapala}\ and\ \citenamefont
  {Moore}(2008)}]{kalapala08}%
  \BibitemOpen
  \bibfield  {author} {\bibinfo {author} {\bibfnamefont {Vamsi}\ \bibnamefont
  {Kalapala}}\ and\ \bibinfo {author} {\bibfnamefont {Cris}\ \bibnamefont
  {Moore}},\ }\bibfield  {title} {\enquote {\bibinfo {title} {{The phase
  transition in exact cover}},}\ }\href@noop {} {\bibfield  {journal} {\bibinfo
   {journal} {Chicago J. Theoret. Comput. Sci.}\ }\textbf {\bibinfo {volume}
  {2008}} (\bibinfo {year} {2008})}\BibitemShut {NoStop}%
\bibitem [{\citenamefont {Islam}\ \emph {et~al.}(2013)\citenamefont {Islam},
  \citenamefont {Senko}, \citenamefont {Campbell}, \citenamefont {Korenblit},
  \citenamefont {Smith}, \citenamefont {Lee}, \citenamefont {Edwards},
  \citenamefont {Wang}, \citenamefont {Freericks},\ and\ \citenamefont
  {Monroe}}]{islam13}%
  \BibitemOpen
  \bibfield  {author} {\bibinfo {author} {\bibfnamefont {R.}~\bibnamefont
  {Islam}}, \bibinfo {author} {\bibfnamefont {C.}~\bibnamefont {Senko}},
  \bibinfo {author} {\bibfnamefont {W.~C.}\ \bibnamefont {Campbell}}, \bibinfo
  {author} {\bibfnamefont {S.}~\bibnamefont {Korenblit}}, \bibinfo {author}
  {\bibfnamefont {J.}~\bibnamefont {Smith}}, \bibinfo {author} {\bibfnamefont
  {A.}~\bibnamefont {Lee}}, \bibinfo {author} {\bibfnamefont {E.~E.}\
  \bibnamefont {Edwards}}, \bibinfo {author} {\bibfnamefont {C.-C.~J.}\
  \bibnamefont {Wang}}, \bibinfo {author} {\bibfnamefont {J.~K.}\ \bibnamefont
  {Freericks}}, \ and\ \bibinfo {author} {\bibfnamefont {C.}~\bibnamefont
  {Monroe}},\ }\bibfield  {title} {\enquote {\bibinfo {title} {{Emergence and
  Frustration of Magnetism with Variable-Range Interactions in a Quantum
  Simulator}},}\ }\href {\doibase 10.1126/science.1232296} {\bibfield
  {journal} {\bibinfo  {journal} {Science}\ }\textbf {\bibinfo {volume}
  {340}},\ \bibinfo {pages} {583--587} (\bibinfo {year} {2013})}\BibitemShut
  {NoStop}%
\bibitem [{\citenamefont {Kirkpatrick}\ \emph {et~al.}(1983)\citenamefont
  {Kirkpatrick}, \citenamefont {Gelatt},\ and\ \citenamefont
  {Vecchi}}]{kirkpatrick83}%
  \BibitemOpen
  \bibfield  {author} {\bibinfo {author} {\bibfnamefont {S.}~\bibnamefont
  {Kirkpatrick}}, \bibinfo {author} {\bibfnamefont {C.~D.}\ \bibnamefont
  {Gelatt}}, \ and\ \bibinfo {author} {\bibfnamefont {M.~P.}\ \bibnamefont
  {Vecchi}},\ }\bibfield  {title} {\enquote {\bibinfo {title} {{Optimization by
  Simulated Annealing}},}\ }\href {\doibase 10.1126/science.220.4598.671}
  {\bibfield  {journal} {\bibinfo  {journal} {Science}\ }\textbf {\bibinfo
  {volume} {220}},\ \bibinfo {pages} {671--680} (\bibinfo {year}
  {1983})}\BibitemShut {NoStop}%
\bibitem [{\citenamefont {Adame}\ and\ \citenamefont
  {McMahon}(2018)}]{adame18}%
  \BibitemOpen
  \bibfield  {author} {\bibinfo {author} {\bibfnamefont {Juan~I.}\ \bibnamefont
  {Adame}}\ and\ \bibinfo {author} {\bibfnamefont {Peter~L.}\ \bibnamefont
  {McMahon}},\ }\bibfield  {title} {\enquote {\bibinfo {title} {{Inhomogeneous
  driving in quantum annealers can result in orders-of-magnitude improvements
  in performance}},}\ }\href@noop {} {\bibfield  {journal} {\bibinfo  {journal}
  {arXiv 1806.11091}\ } (\bibinfo {year} {2018})}\BibitemShut {NoStop}%
\bibitem [{\citenamefont {Hauke}\ \emph {et~al.}(2015)\citenamefont {Hauke},
  \citenamefont {Bonnes}, \citenamefont {Heyl},\ and\ \citenamefont
  {Lechner}}]{hauke15}%
  \BibitemOpen
  \bibfield  {author} {\bibinfo {author} {\bibfnamefont {Philipp}\ \bibnamefont
  {Hauke}}, \bibinfo {author} {\bibfnamefont {Lars}\ \bibnamefont {Bonnes}},
  \bibinfo {author} {\bibfnamefont {Markus}\ \bibnamefont {Heyl}}, \ and\
  \bibinfo {author} {\bibfnamefont {Wolfgang}\ \bibnamefont {Lechner}},\
  }\bibfield  {title} {\enquote {\bibinfo {title} {{Probing entanglement in
  adiabatic quantum optimization with trapped ions}},}\ }\href {\doibase
  10.3389/fphy.2015.00021} {\bibfield  {journal} {\bibinfo  {journal} {Front.
  Phys.}\ }\textbf {\bibinfo {volume} {3}},\ \bibinfo {pages} {21} (\bibinfo
  {year} {2015})}\BibitemShut {NoStop}%
\bibitem [{\citenamefont {Gra{\ss}}\ \emph {et~al.}(2016)\citenamefont
  {Gra{\ss}}, \citenamefont {Ravent{\'o}s}, \citenamefont {Juli{\'a}-D{\'i}az},
  \citenamefont {Gogolin},\ and\ \citenamefont {Lewenstein}}]{grass16}%
  \BibitemOpen
  \bibfield  {author} {\bibinfo {author} {\bibfnamefont {Tobias}\ \bibnamefont
  {Gra{\ss}}}, \bibinfo {author} {\bibfnamefont {David}\ \bibnamefont
  {Ravent{\'o}s}}, \bibinfo {author} {\bibfnamefont {Bruno}\ \bibnamefont
  {Juli{\'a}-D{\'i}az}}, \bibinfo {author} {\bibfnamefont {Christian}\
  \bibnamefont {Gogolin}}, \ and\ \bibinfo {author} {\bibfnamefont {Maciej}\
  \bibnamefont {Lewenstein}},\ }\bibfield  {title} {\enquote {\bibinfo {title}
  {{Quantum annealing for the number-partitioning problem using a tunable spin
  glass of ions}},}\ }\href {\doibase 10.1038/ncomms11524;;;;;;;;;;;
  10.1038/ncomms11524} {\bibfield  {journal} {\bibinfo  {journal} {Nat.
  Commun.}\ }\textbf {\bibinfo {volume} {7}},\ \bibinfo {pages} {11524}
  (\bibinfo {year} {2016})}\BibitemShut {NoStop}%
\bibitem [{\citenamefont {Glaetzle}\ \emph {et~al.}(2017)\citenamefont
  {Glaetzle}, \citenamefont {van Bijnen}, \citenamefont {Zoller},\ and\
  \citenamefont {Lechner}}]{glaetzle17}%
  \BibitemOpen
  \bibfield  {author} {\bibinfo {author} {\bibfnamefont {A.~W.}\ \bibnamefont
  {Glaetzle}}, \bibinfo {author} {\bibfnamefont {R.~M.~W.}\ \bibnamefont {van
  Bijnen}}, \bibinfo {author} {\bibfnamefont {P.}~\bibnamefont {Zoller}}, \
  and\ \bibinfo {author} {\bibfnamefont {W.}~\bibnamefont {Lechner}},\
  }\bibfield  {title} {\enquote {\bibinfo {title} {{A coherent quantum annealer
  with Rydberg atoms}},}\ }\href {\doibase 10.1038/ncomms15813;;;;;
  10.1038/ncomms15813} {\bibfield  {journal} {\bibinfo  {journal} {Nat.
  Commun.}\ }\textbf {\bibinfo {volume} {8}},\ \bibinfo {pages} {15813}
  (\bibinfo {year} {2017})}\BibitemShut {NoStop}%
\bibitem [{\citenamefont {Torggler}\ \emph {et~al.}(2017)\citenamefont
  {Torggler}, \citenamefont {Kr{\"a}mer},\ and\ \citenamefont
  {Ritsch}}]{torggler17}%
  \BibitemOpen
  \bibfield  {author} {\bibinfo {author} {\bibfnamefont {Valentin}\
  \bibnamefont {Torggler}}, \bibinfo {author} {\bibfnamefont {Sebastian}\
  \bibnamefont {Kr{\"a}mer}}, \ and\ \bibinfo {author} {\bibfnamefont {Helmut}\
  \bibnamefont {Ritsch}},\ }\bibfield  {title} {\enquote {\bibinfo {title}
  {{Quantum annealing with ultracold atoms in a multimode optical
  resonator}},}\ }\href {\doibase 10.1103/PhysRevA.95.032310} {\bibfield
  {journal} {\bibinfo  {journal} {Phys. Rev. A}\ }\textbf {\bibinfo {volume}
  {95}},\ \bibinfo {pages} {032310} (\bibinfo {year} {2017})}\BibitemShut
  {NoStop}%
\bibitem [{\citenamefont {Richerme}\ \emph {et~al.}(2013)\citenamefont
  {Richerme}, \citenamefont {Senko}, \citenamefont {Korenblit}, \citenamefont
  {Smith}, \citenamefont {Lee}, \citenamefont {Islam}, \citenamefont
  {Campbell},\ and\ \citenamefont {Monroe}}]{richerme13}%
  \BibitemOpen
  \bibfield  {author} {\bibinfo {author} {\bibfnamefont {P.}~\bibnamefont
  {Richerme}}, \bibinfo {author} {\bibfnamefont {C.}~\bibnamefont {Senko}},
  \bibinfo {author} {\bibfnamefont {S.}~\bibnamefont {Korenblit}}, \bibinfo
  {author} {\bibfnamefont {J.}~\bibnamefont {Smith}}, \bibinfo {author}
  {\bibfnamefont {A.}~\bibnamefont {Lee}}, \bibinfo {author} {\bibfnamefont
  {R.}~\bibnamefont {Islam}}, \bibinfo {author} {\bibfnamefont {W.~C.}\
  \bibnamefont {Campbell}}, \ and\ \bibinfo {author} {\bibfnamefont
  {C.}~\bibnamefont {Monroe}},\ }\bibfield  {title} {\enquote {\bibinfo {title}
  {{Quantum Catalysis of Magnetic Phase Transitions in a Quantum Simulator}},}\
  }\href {\doibase 10.1103/PhysRevLett.111.100506} {\bibfield  {journal}
  {\bibinfo  {journal} {Phys. Rev. Lett.}\ }\textbf {\bibinfo {volume} {111}},\
  \bibinfo {pages} {100506} (\bibinfo {year} {2013})}\BibitemShut {NoStop}%
\end{thebibliography}

\end{document}